\documentclass[amsmath,amssymb,aps]{revtex4-2}

\usepackage{amsmath,amssymb,array,epsfig,graphicx,graphics,color,float,caption}

\newcommand{\enquote}[1]{``#1''}

\begin{document}

\title{Interacting ring-Airy beams in nonlinear media}

\author{Charles W. Robson}
\email{charles.robson@tuni.fi}
\author{Marco Ornigotti}
\affiliation{Laboratory of Photonics, Physics Unit, Tampere University, Tampere, FI-33720 Finland}



\begin{abstract}
The interactions between copropagating ring-Airy beams in a (2+1)-dimensional Kerr medium are numerically investigated for the first time. It is shown that two overlapping ring-Airy beams in such a medium produce controllable regions of very low intensity during propagation, the geometry of which can be manipulated by the tuning of initial beam parameters. This may prove useful for future optical tweezing applications in the nonlinear regime.
\end{abstract}

\maketitle

\vspace{-2mm}

\section{Introduction}

Since the first experimental observation of finite-energy Airy optical beams was reported \cite{Airy_exp_PRL} in 2007, there has been a wealth of research on these non-diffracting, freely-accelerating beams in both the linear and nonlinear regimes \cite{Peng1,Peng2,Papa_natcomms,Fedorov,Airy_book_chapter,Christo1,Christo2,Zhang,Baumgartl}. Originally studied in the context of quantum mechanics \cite{Berry,Comment_to_Berry}, Airy wave packets have entered the field of optics due to the equivalent mathematical descriptions of quantum particles (the Schr{\" o}dinger equation) and optical waves via the paraxial equation.

Several variations of the original Airy beam have been studied in recent years \cite{Airy_Bessel,Bandres,Cosh_Airy,Cosh_Airy2,Airy_vortex,Deng,Zhong,Ince}, one being a radially-symmetric, exponentially-adopized version, known as the ring-Airy beam \cite{Christo1,Christo2}. This beam has several intriguing features, such as the ability to rapidly and strongly autofocus \cite{Christo1}. Although the interaction properties of several different variants of the original Airy beam in nonlinear media have been studied \cite{Soliton_pair_gen,Rudnick,Chen,Zhou}, surprisingly, the interactions between ring-Airy beams in these media do not seem to have been treated.

In this Letter, we investigate, for the first time to our knowledge, the dynamics of two overlapping ring-Airy beams (RABs) in a local Kerr medium. The interactions are shown to possess several interesting features, such as spatially-extended zero-intensity regions and a high degree of controllability of these regions: these are potentially useful for future applications in optical tweezing in nonlinear media.

It is known \cite{Peng1,Peng2,Papa_natcomms} that a single RAB self-focuses in a Kerr medium and that it can form a local region of zero (or very low) intensity during propagation, potentially useful for particle trapping. Previously \cite{Peng1}, the axial location of this region of low intensity -- which we shall call a \enquote{node} from hereon -- has only been shown to be controllable via a tuning of the nonlocality of the Kerr medium in which the beam propagates. In the present work, we demonstrate that this node position can in fact be controlled without the need for such nonlocal fine-tuning: both for the case of two overlapping RABs as well as of propagation of a single RAB.

We focus on two different cases of interacting RABs in this Letter: firstly, two overlapping RABs with zero initial phase difference and, secondly, two beams possessing an initial phase difference of $\pi$; the effect of this initial phase difference is shown to be considerable.

\section{Method}

All results and figures were generated using numerical simulations which solved the (2+1)-dimensional nonlinear Schr\"{o}dinger (2+1 NLS) equation using a split-step Fourier method.

We developed a MATLAB\textsuperscript{\tiny\textregistered} code to simulate the 2+1 NLS equation: this code was built upon that given in Appendix B of Ref. \cite{Agrawal}, which we adapted and generalised for our purposes.

\section{Results}

\subsection{Zero phase difference at input}

Ring-Airy beams are radially-symmetric field configurations built from the Airy function, first advanced \cite{Christo1} in 2010, and seen experimentally \cite{Christo2} the following year, and can be expressed as \cite{Peng1,Peng2,Christo1}
\begin{equation} \label{Airy_ring}
u\left( x,y,z=0 \right) = A.\mathrm{Ai}\left( r_{0}-\sqrt{x^2 + y^2} \right) \mathrm{exp}\left( a \left( r_{0}-\sqrt{x^2 + y^2} \right) \right) ,
\end{equation}
with $\mathrm{Ai}$ the Airy function, $A$ the amplitude of the field, $a$ a decay parameter (to ensure finite energy), and $r_{0}$ a radial parameter controlling the form of the field (specifically, the radius of the first bright ring of the beam in the transverse plane).

The reason that $z$ is set to zero in Eq. (\ref{Airy_ring}) is that we would like to simulate the evolution of the RAB, taken as an initial condition, using the \enquote{defocusing} 2+1 NLS equation \cite{Sulems,Agrawal2}:
\begin{equation} \label{NLS}
i \frac{\partial u}{\partial z} - \frac{1}{2}\frac{\partial^2 u}{\partial x^2} - \frac{1}{2}\frac{\partial^2 u}{\partial y^2} + |u|^2 u = 0.
\end{equation}

It has recently been shown \cite{Peng1,Peng2} that a ring-Airy input will exhibit interesting dynamics in a medium described by the 2+1 NLS equation, including a \enquote{double focus} as well as a low-intensity axial region, the position of the latter being controllable by altering the degree of nonlocality of the medium; these features have been argued \cite{Peng1,Peng2} to be potentially useful for particle trapping in nonlinear media, i.e. for optical tweezing in the nonlinear regime.

In this Letter, we show that a spatially-controllable low-intensity region can exist without the need for tuning of the degree of nonlocality in a Kerr medium, for the case of two overlapping RABs as well as that of a single propagating RAB: the former case manifests a richer field evolution and can be tuned to give different node geometries: for example, a nodal ``corridor" can form when the two beams are initially completely out of phase.

The two overlapping RABs we choose as input have the following form
\begin{equation} \label{Two_RABs}
\begin{split}
u\left( x,y,z=0 \right) =& A.\mathrm{Ai} \left( r_{0}-\sqrt{(x-0.8)^2 + y^2} \right) \mathrm{exp}\left( a \left( r_{0}-\sqrt{(x-0.8)^2 + y^2} \right) \right) \\
&+ A.\mathrm{Ai} \left( r_{0}-\sqrt{(x+0.8)^2 + y^2} \right) \mathrm{exp}\left( a \left( r_{0}-\sqrt{(x+0.8)^2 + y^2} \right) \right) .
\end{split}
\end{equation}
As can be seen, each RAB is of the same form as (\ref{Airy_ring}), but shifted transversely: one RAB is shifted negatively along the $x$-axis and the other positively.

\begin{figure}
  \centering
  \includegraphics[width=0.5\linewidth]{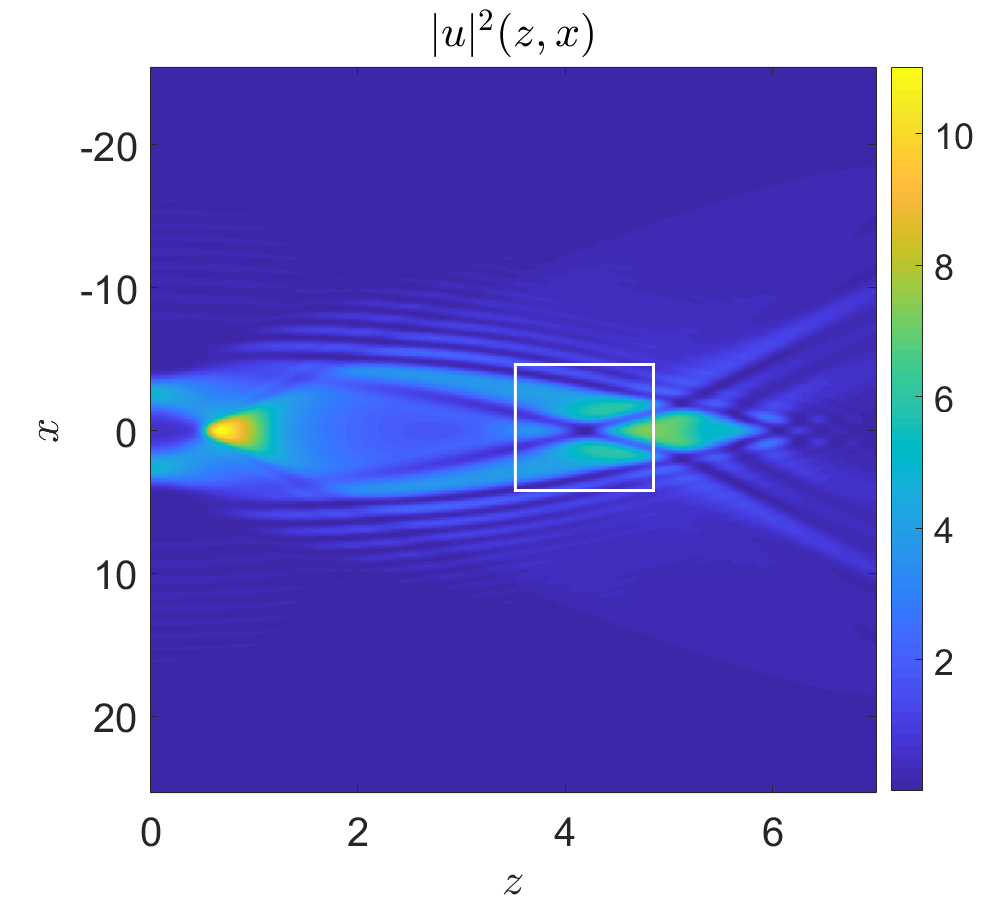}
  \caption{Intensity evolution of the field given by Eq. (\ref{Two_RABs}) in the ($z$,$x$) plane. The X-shaped low-intensity region, centred on the node, is highlighted. (Intensities are in arbitrary units.)}
  \label{fig:overlap}
\end{figure}

\begin{figure}
  \centering
  \includegraphics[width=0.5\linewidth]{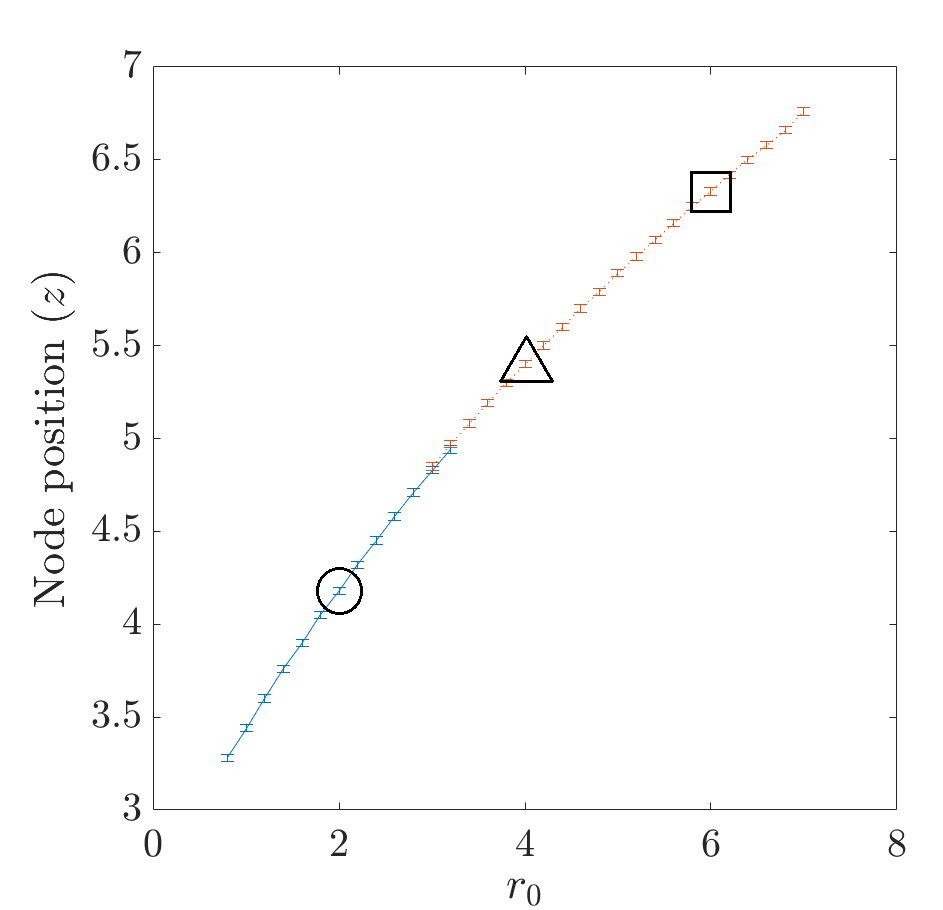}
  \caption{Numerical results showing the dependence of node position on the value of $r_{0}$. The blue, solid fit line is for $a=0.1$, and the red, dotted fit line is for $a=0.2$.}
  \label{fig:combined_trend}
\end{figure}

\begin{figure}
  \centering
  \includegraphics[width=0.5\linewidth]{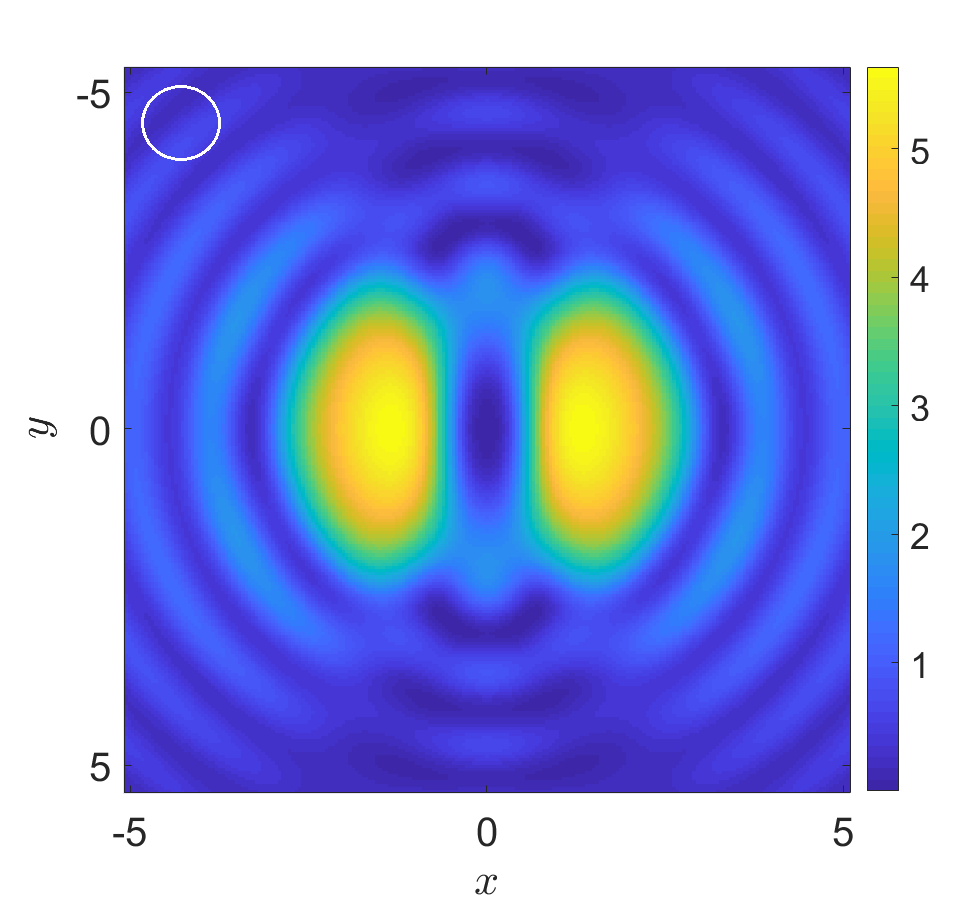}
  \caption{A cross section showing the intensity in the ($x$,$y$) plane at the $z$ position indicated by the circle in Fig. \ref{fig:combined_trend}.}
  \label{fig:cross_sect_1}
\end{figure}

\begin{figure}
  \centering
  \includegraphics[width=0.5\linewidth]{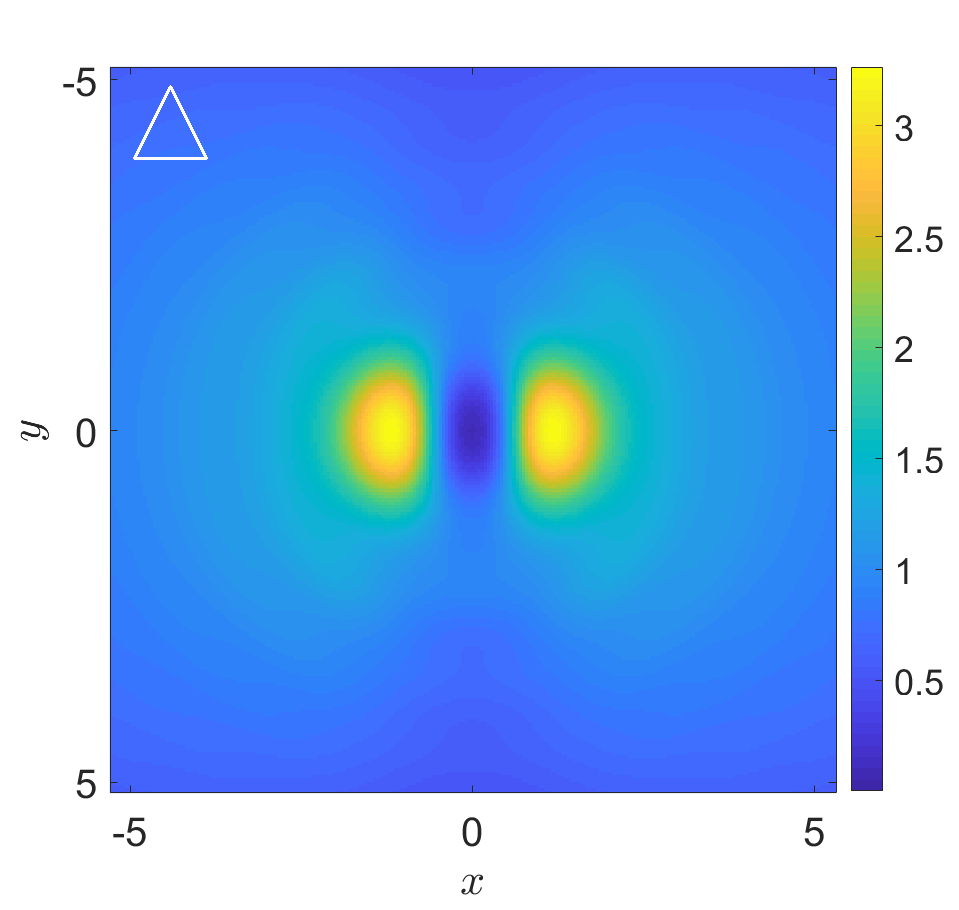}
  \caption{A cross section showing the intensity in the ($x$,$y$) plane at the $z$ position indicated by the triangle in Fig. \ref{fig:combined_trend}.}
  \label{fig:cross_sect_2}
\end{figure}

\begin{figure}
  \centering
  \includegraphics[width=0.5\linewidth]{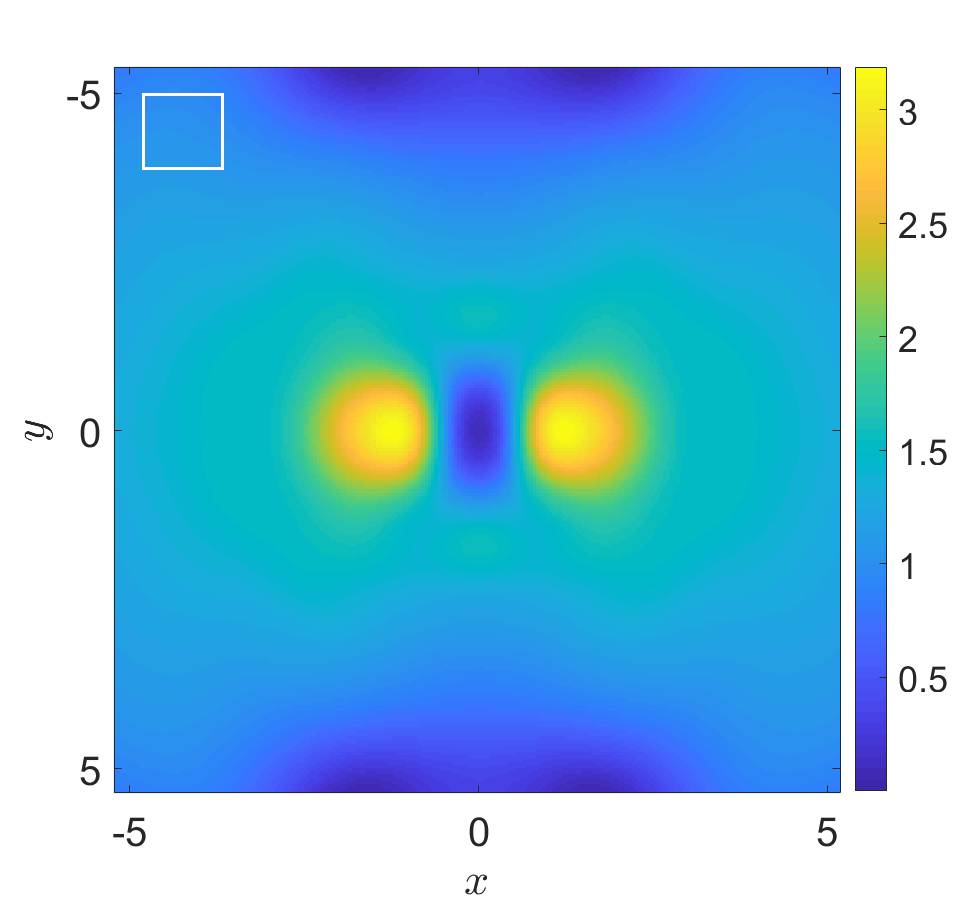}
  \caption{A cross section showing the intensity in the ($x$,$y$) plane at the $z$ position indicated by the square in Fig. \ref{fig:combined_trend}.}
  \label{fig:cross_sect_3}
\end{figure}

\begin{figure}
  \centering
  \includegraphics[width=0.5\linewidth]{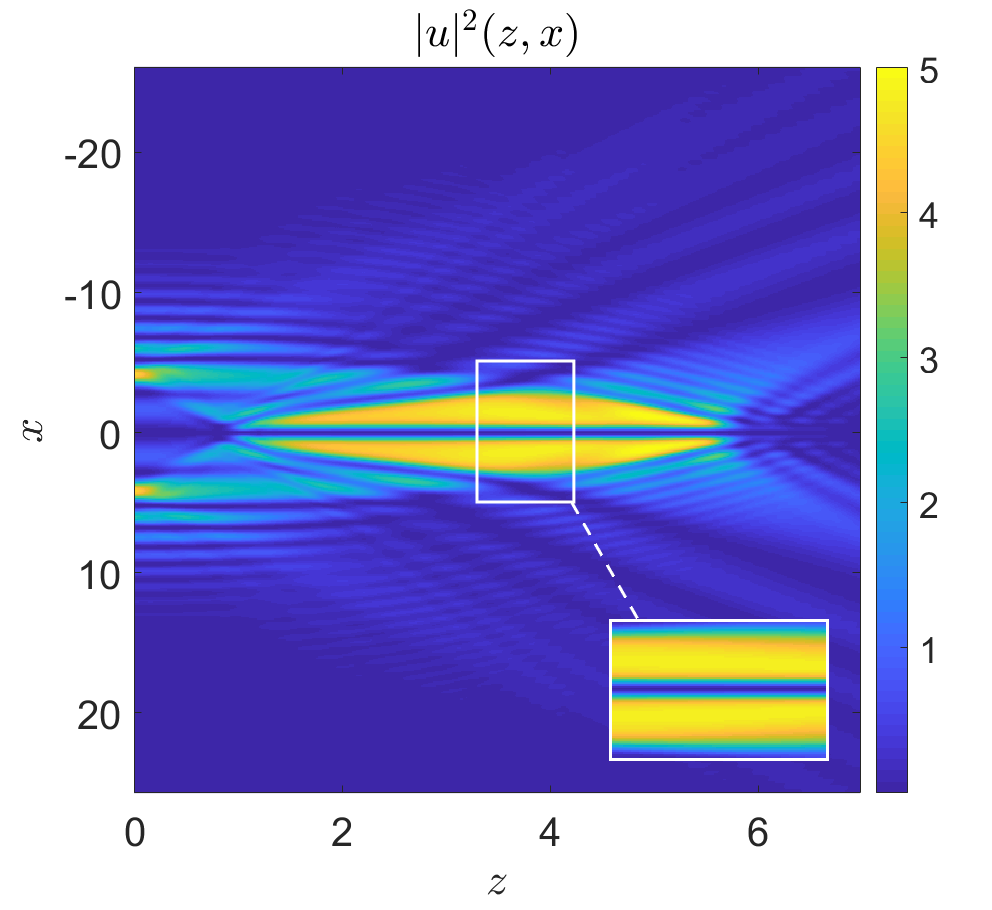}
  \caption{Evolution of two overlapping phase-shifted RABs with a part of the low-intensity region highlighted.}
  \label{fig:phase_diff_with_box}
\end{figure}

Fig. \ref{fig:overlap} shows how the intensity of this input varies in the ($z$,$x$) plane during propagation in a Kerr medium when $r_{0}=2$ (here $y=0$ and the input field parameters are set to $a=0.1$ and $A=3$). The white box in the figure highlights the aforementioned node. It is to be noted that, for these input parameter values, if just one RAB is used as input -- i.e. just one of the terms from the RHS of Eq. (\ref{Two_RABs}) -- then there is no nodal point.

By varying the radial parameter $r_{0}$, whilst keeping all other parameters fixed, the position of the node along $z$ shifts in a remarkably ordered way, as shown in Fig. \ref{fig:combined_trend}: the trend is monotonic and almost linear. When $r_{0}$ increases to a threshold value of $r_{0} {\sim} 3$, the low-intensity region starts to bifurcate for the parameter choice $a=0.1$. By increasing its value to $a=0.2$, whilst keeping the field amplitude the same ($A=3$), splitting of the low-intensity regions can be avoided. Figs. \ref{fig:cross_sect_1}, \ref{fig:cross_sect_2}, and \ref{fig:cross_sect_3} show field intensities in the transverse plane at the propagation points along $z$ signified in Fig. \ref{fig:combined_trend}.

Controllable nodal regions have recently been numerically predicted for a single RAB input in \emph{nonlocal} nonlinear media \cite{Peng1, Peng2}; in contrast, the results presented in this Letter do not require any nonlocality of the medium (and hence no fine-tuning of the nonlocality): the control emerges simply from changing the initial parameters of the beam. It should be noted that local control of nodal positions is also seen in our numerical results for the evolution of a \emph{single} RAB, again via the tuning of $r_{0}$ values: for a single beam, a node is present only when its initial amplitude exceeds $A{\sim}10$ (a similar result was shown in Fig. 1c of Ref. \cite{Peng1}). The main advantage of using two beams, as opposed to one, is the extra degrees of freedom introduced into the system which could potentially aid future optical tweezing applications where controllability of fields is paramount. In the following section, we look at one example due to these extra degrees of freedom: an initial $\pi$-phase difference between the beams leading to a vastly different field evolution.

\subsection{Non-zero phase difference at input}

The case is now treated in which two input RABs have an initial phase difference of $\Delta \phi_{0}=\pi$, leading to a very different intensity evolution. The new phase-shifted form of the initial field is
\begin{equation} \label{Two_RABs_phase_shifted}
\begin{split}
u \left( x,y,z=0 \right) =& A.\mathrm{Ai} \left( r_{0}-\sqrt{(x-0.8)^2 + y^2} \right) \mathrm{exp} \left( a \left( r_{0}-\sqrt{(x-0.8)^2 + y^2} \right) \right) \rm{exp} \left( i\pi \right) \\
&+ A.\mathrm{Ai}\left( r_{0}-\sqrt{(x+0.8)^2 + y^2} \right) \mathrm{exp} \left( a \left( r_{0}-\sqrt{(x+0.8)^2 + y^2} \right) \right) ;
\end{split}
\end{equation}
cf. Eq. (\ref{Two_RABs}).

A striking change in field evolution occurs when this initial phase difference is incorporated. The two RABs no longer create a nodal point somewhere along the longitudinal direction, but, instead, create a nodal corridor: a hollow, extended region around the centre of the beam.

\begin{figure}
    \centering
    \includegraphics[width=0.5\linewidth]{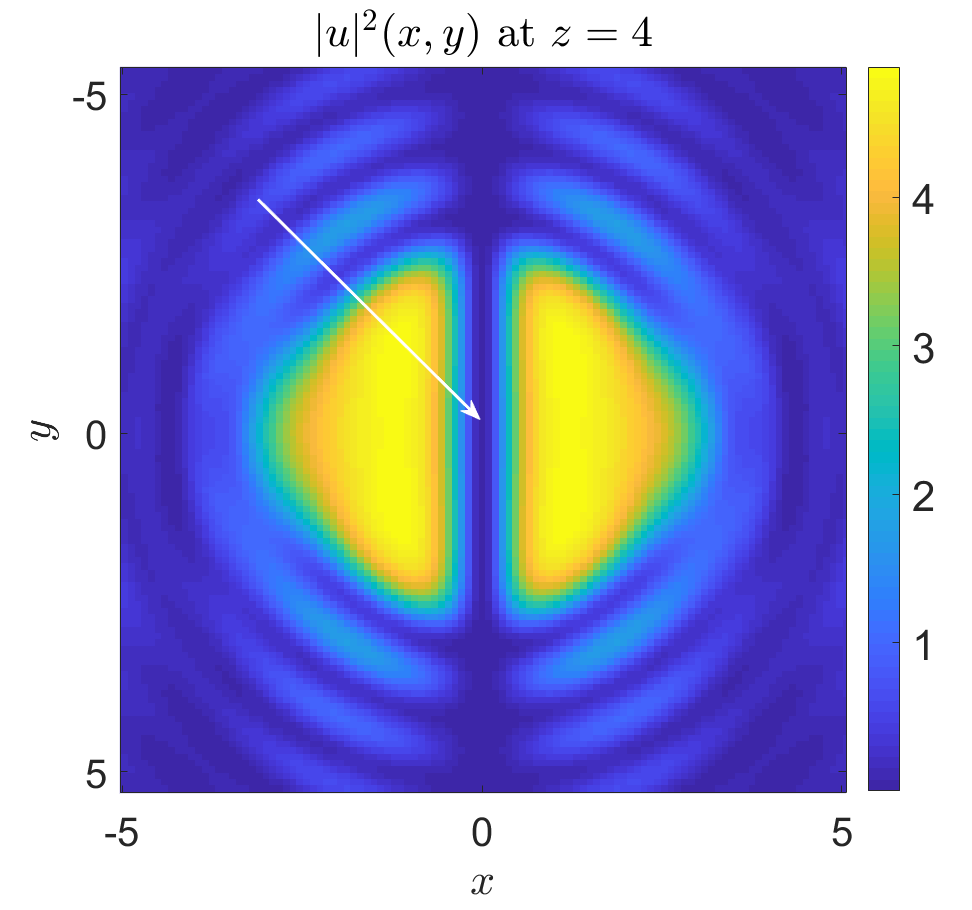}
    \caption{A cross section showing the intensity of the field given by Eq. (\ref{Two_RABs_phase_shifted}) after propagating in the medium to $z=4$. The white arrow highlights a transverse slice of the low-intensity corridor.}
    \label{fig:phase_shift_cross_sect}
\end{figure}

Fixing the parameters as $a=0.1$, $r_{0}=2$ and $A=3$ and evolving input field (\ref{Two_RABs_phase_shifted}) using the NLS equation (\ref{NLS}) results in an extended and low-intensity region emerging in the ($z$, $x$) plane, as shown in Fig. \ref{fig:phase_diff_with_box}. This central, low-intensity region has, in fact, practically \emph{zero intensity}, and is surrounded by high-intensity regions. One potential application of this large intensity gradient could be the manipulation of particles into a planar geometry. See Fig. \ref{fig:phase_shift_cross_sect} for a cross section of the nodal corridor in the transverse plane.




\section{Discussion}

The controllable low-intensity regions illustrated in this Letter may prove useful for optical tweezing in nonlinear media. The regions surrounding the nodes have high intensities, relative to the rest of the field, suggesting there may be a strong gradient force on particles if they were to be present in the system: one appropriate medium to test this suggestion could be a dye-doped liquid crystal, which has been shown to sustain a large negative nonlinearity \cite{Khoo}. An analysis of the radiation forces \cite{Harada} acting on a dielectric particle due to these overlapping RABs will be the subject of a future publication.

Apart from the aforementioned practical application, the interactions between ring-Airy beams in local Kerr media is interesting in and of itself, especially as it seems, surprisingly, to have been neglected in the literature. It may be fruitful for future studies to investigate the interactions between many (more than two) RABs, as well as to look into how the added degrees of freedom that we have described can be exploited to create new light geometries during propagation, for example by copropagating beams of different amplitudes and with $\Delta \phi_{0} \neq \pi$.

\section*{Disclosures}

The authors declare no conflicts of interest.

\section{Conclusion}

In this Letter, we have investigated numerically the interactions between ring-Airy beams in a local Kerr medium. The results show that the evolution of two overlapping ring-Airy beams produces regions of low intensity at specific points along the propagation axis, and the addition of an initial phase difference between the overlapping beams allows for richer low-intensity geometries. These nodes could be well-suited to optical tweezing in a nonlinear medium due to the surrounding presence of high-intensity regions, creating a strong field gradient. We have also demonstrated for the first time that that there is no requirement to include nonlocality in a Kerr medium in order to control the position of nodal regions: a simple adjustment of initial beam parameters suffices.

\section{Acknowledgments}

This work is part of the Academy of Finland Flagship Programme, Photonics Research and Innovation (PREIN), decision 320165.



\end{document}